\pdfoutput=1
\documentclass[twocolumn,floatfix,prl,aps,reprint]{revtex4-1}
\usepackage{float}
\usepackage{graphicx}
\usepackage{amsmath}
\usepackage{bm}
\usepackage{color}
\usepackage{amssymb}
\usepackage{xspace}
\usepackage{epstopdf}
\usepackage{dcolumn}
\usepackage{longtable}
\usepackage{cancel}

\usepackage[colorlinks=true, letterpaper=true, pdfstartview=FitV, linkcolor=blue, citecolor=blue, urlcolor=blue]{hyperref}
\begin{document}
\title{Flavor Symmetry and Ferroelectric Nematics in Transition Metal Dichalcogenides}
\author{Patrick Cheung}\author{Zhi-qiang Bao}\author{Fan Zhang}\email{zhang@utdallas.edu}
\affiliation{Department of Physics, The University of Texas at Dallas, Richardson, TX 75080, USA}
\begin{abstract}
Recent magneto-transport experiments have provided compelling evidence for the presence of 
an energetically isolated threefold $Q$-valley degeneracy in few-layer transition metal dichalcogenides.   
We study the flavor SU(3) symmetry breaking when each Landau level triplet is one-third filled or empty and 
predict that a pure flavor nematic phase and a flavorless charge-density-wave phase 
will occur respectively below and above a critical magnetic field.
Surprisingly, electrons carry flavor-dependent electric dipole moments even at zero magnetic field, 
rendering the nematics ferroelectric, allowing electric-field manipulation of the flavors, 
and leading to the concept of flavortronics. 
\end{abstract}
\maketitle

{\color{blue}\indent{\it Introduction.}}---Spontaneous symmetry breaking associated with the internal degrees 
of freedom of electrons plays a central role in determining the electromagnetic phenomena in materials. 
Prime examples are charge U(1) symmetry breaking in superconductors and spin SU(2) symmetry breaking in ferromagnets. 
SU(2) symmetries are also innate to double-layer and dual-valley systems 
such as few-layer graphene and transition metal dichalcogenides (TMDs)~\cite{graphene,intro1,intro2}.
However, flavor SU(3) symmetry appears to be rare in electron systems, 
although historically it provides a foundation for the quark model~\cite{hep}. 
This fact prevents the emergence of novel electromagnetic phenomena that arise from the flavor symmetry or its breaking. 
Recently, experiments~\cite{Q1,Q2,Q3,Q4,Q5} have provided compelling evidence  
for the presence of an energetically isolated threefold $Q$-valley degeneracy in few-layer TMDs
including MoS$_2$, MoSe$_2$, WS$_2$, and WSe$_2$.
From this perspective, the TMDs offer an unprecedented platform for exploring  
the flavor SU(3) symmetry of electrons. 
(The surface of Bi or SnTe~\cite{Ong,Zhu,BiLL,SnTe,SnTe111,XiaoLi} also hosts triply degenerate valleys,
yet they coexist with other ones that cannot be removed from Fermi energy.) 

\begin{figure}[t!]
\includegraphics[width=1.0\columnwidth]{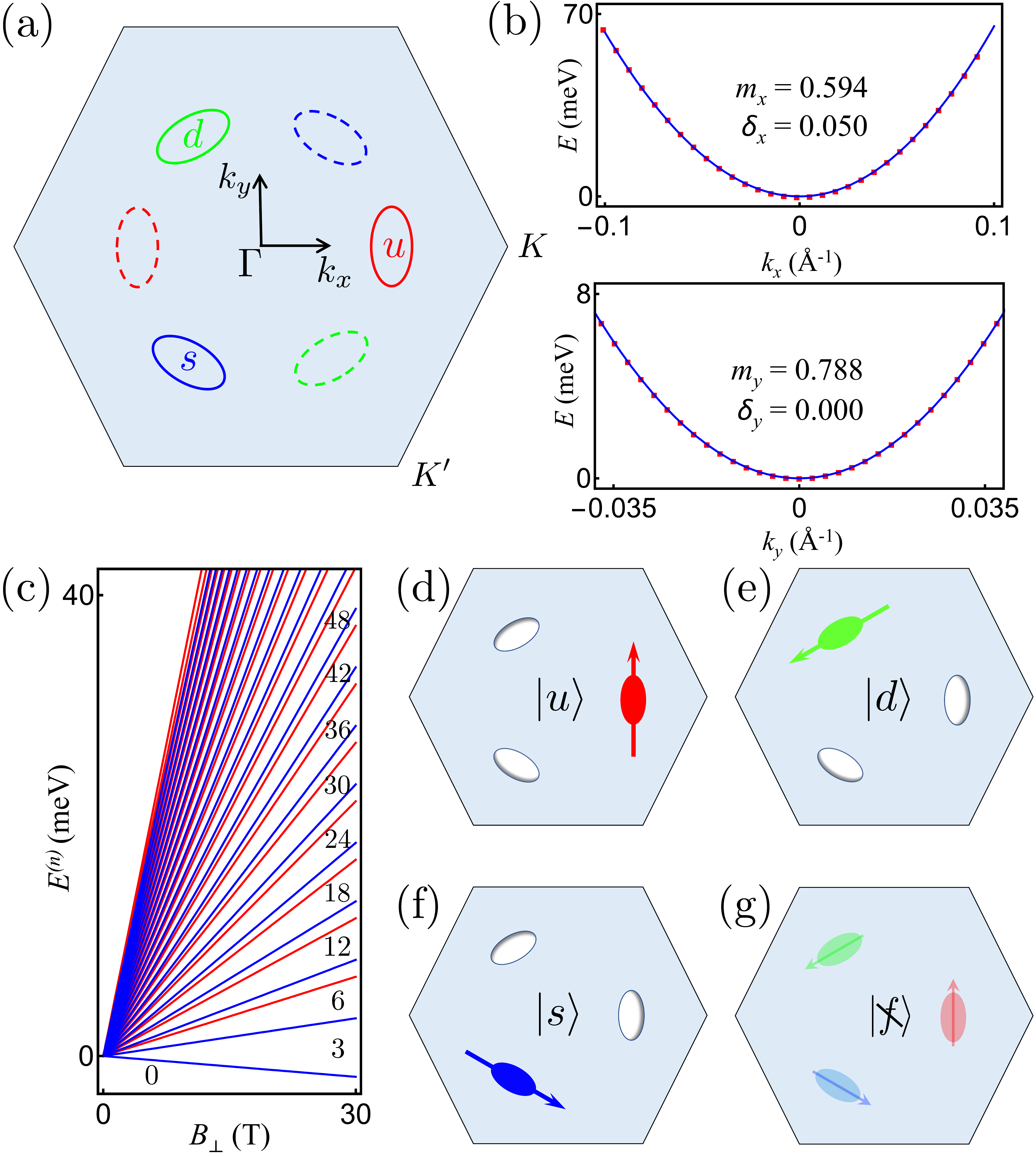}
\caption{(a) Schematic of the first BZ of few-layer TMDs. 
The conduction band minima are located at the six $Q$ (solid) and $Q'$ (dashed) valleys. 
When both $\mathcal{T}$ and $\mathcal{P}$ symmetries are broken, 
each $Q$-valley LL triplet ($u,d,s$) has an emergent flavor SU(3) symmetry, 
analogous to the quark model. 
(b) Band parameter values in Eq.~(\ref{Ham}) for trilayer MoS$_2$ 
extracted from the first-principles calculations (squares).
(c) Single-particle LL structure~\cite{LLnote} of (b) with $g^*=5$. 
The $Q$ (blue) and $Q'$ (red) LL triplets have opposite spins and Zeeman split,
yielding quantum Hall states at $\nu=3N$ ($N\!\in\!{\mathbb N}$).
(d)-(g) Schematics of interaction-driven broken flavor symmetry states at $\nu=3N-2$. 
The pure flavor states, {i.e.}, $|u\rangle, |d\rangle$, and $|s\rangle$, are ferroelectric nematics.
The flavorless states, {e.g.}, $|\bcancel{f}\rangle=(|u\rangle + |d\rangle + |s\rangle)/\sqrt{3}$, 
are incommensurate CDW states.
The color fillings indicate particle populations in the highest triplet, 
and the arrows depict electric dipole moments.}
\label{fig1}
\end{figure}

At the conduction-band edge of a few-layer TMD, 
there are six inequivalent $Q/Q'$ valleys inside the first Brillouin zone (BZ), as depicted in Fig~\ref{fig1}a. 
The three $Q$ valleys are related by a $\mathcal{C}_3$ symmetry, 
and the $Q$ and $Q'$ valleys are related by time-reversal ($\mathcal{T}$) and inversion ($\mathcal{P}$) symmetries. 
In an odd-layer system, because of the intrinsic $\mathcal{P}$ asymmetry and strong spin-orbit couplings, 
the bands are spin split. 
The $z$-component of spin is a good quantum number due to the mirror symmetry 
with respect to the middle metal layer. 
Thus, the $Q$-valley bands exhibit both spin splitting and spin-valley locking. 
In an even-layer system, the restored $\mathcal{P}$ symmetry and the $\mathcal{T}$ symmetry 
leads to Kramers degeneracy for all the states. As a result, the bands are spin degenerate in each $Q$ valley.
Such universal even-odd layer-dependent valley degeneracies have been clearly observed    
in the Shubnikov-de Hass oscillations~\cite{Q1,Q2,Q3}. 
At small magnetic fields and low electron densities, 
the oscillations display a $6$ ($12$) fold Landau level (LL) degeneracy 
in the odd- (even-) layer systems due to the $\mathcal{C}_3 \times \mathcal{T}$ 
($\mathcal{C}_3 \times \mathcal{T} \times \mathcal{P}$) symmetry. At large fields, 
the inter- (intra-) valley Zeeman effect reduces the degeneracy to $3$ ($6$). 

Therefore, at magnetic fields $\gtrsim5$~T and electron densities $\lesssim10^{13}$~cm$^{-2}$~\cite{Q1}, 
all the TMD odd-layers thicker than bilayer and the even-layers with an interlayer electric field 
(to break the $\mathcal{P}$ symmetry) exhibit LL triplets.
Because of the similarity to the quark model, and to distinguish from the $Q/Q'$ valley pseudospin, 
it is proper to label the three degenerate $Q$ valleys of orientations $\theta_f\in\{0, 2\pi/3, 4\pi/3\}$
by their {\em flavors} $f\in\{u, d, s\}$ as in Fig.~\ref{fig1}a. 
In this unique platform, we examine the interplay between the flavor SU(3) symmetry 
and electron-electron interactions, by focusing on the case in which the highest 
LL triplet is $1/3$ filled or empty to produce an integer quantum Hall effect. 
When only the flavor-conserving long-range interactions are retained, the $\mathcal{C}_3$ symmetry is spontaneously broken 
to yield a nematic state in which all the electrons have the same flavor. 
When the much weaker flavor-mixing short-range interactions are taken into account, 
the translational symmetry is lowered beyond a critical magnetic field to create  
incommensurate charge-density-wave (CDW) states that establish coherence among three flavors.   
Remarkably, the large effective mass of TMD electrons substantially magnifies the interaction effects, 
and the lack of inversion centers of $Q$-valleys renders the nematics ferroelectric.
This electronic ferroelectricity~\cite{F00,F01,F02} differs fundamentally from the structural one~\cite{F1,F2,F3,F4}.
Interestingly, an in-plane electric field couples to the electron flavor through its flavor-dependent 
electric dipole moment that survives even at zero magnetic field, 
thereby tuning the ferroelectric nematics, CDW states, and their competition.

{\color{blue}\indent{\it Model.}}---Near the energy minima (${\bm k}=0$) of the lowest conduction 
sub-band of a TMD few-layer, the most general flavor-dependent $Q$-valley Hamiltonians read
\begin{equation}
\begin{aligned}
H_u({\bm k}) = \frac{\hbar^2k_{x}^2}{2m_x} + \frac{\hbar^2k_{y}^2}{2m_y} + \frac{\delta_x a\hbar^2 k_x^3}{m_e},
\end{aligned}\label{Ham}
\end{equation}
$H_d=\mathcal{C}_3H_u\mathcal{C}_3^{-1}$,  
and $H_s=\mathcal{C}_3^{-1}H_u\mathcal{C}_3$. 
Here $m_e$ is the bare electron mass, and $a$ is the lattice constant, {e.g.}, $3.160$~\AA~for MoS$_2$.
The energy dispersions are nearly quadratic, with different effective masses $m_x$ and $m_y$ 
along and perpendicular to the $\Gamma K$ direction, respectively. 
While each $Q$ valley is symmetric with respect to its axis $\Gamma K$, there is an asymmetry along it, 
and $\delta_x$ characterizes this weak distortion in Eq.~(\ref{Ham}).
These band parameters can be extracted from first-principles calculations,  
and $\sqrt{m_xm_y}$ can alternatively be obtained from temperature-dependent SdH oscillations~\cite{Q1,Q2,Q3}. 
Note that their values crucially depend on the layer number. For trilayer MoS$_2$, 
our first-principles calculations yield $m_x = 0.594 m_e$, $m_y = 0.788 m_e$, and $\delta_x=0.050$, 
as fitted in Fig.~\ref{fig1}b.
We ignore the tiny $\delta_x$ when deriving the LLs and the broken symmetry states 
yet treat it perturbatively when computing the electric polarization. 

Under a uniform perpendicular magnetic field $B_{\perp}$, $\hbar\bm{k}$ are replaced by 
$\bm{\pi} = \hbar\bm{k} + e\bm{A}$ with $\bm{A} = (0,-B_{\perp}x)$ in the Landau gauge. 
Since $[\pi_x,\pi_y]=i\hbar^2/\ell^2$ with $\ell = \sqrt{\hbar/eB_{\perp}}$ being the magnetic length, 
we construct a set of raising operators
$a^\dagger_f = (\ell/\sqrt{2}\hbar)(\alpha_f \pi_x - i\beta_f\pi_y)$, 
where $\alpha_f = \eta\cos\theta_f +  i\eta^{-1}\sin\theta_f$, 
$\beta_f = \eta^{-1}\cos\theta_f + i\eta\sin\theta_f$, and $\eta = (m_y/m_x)^{1/4}$. 
This flavor-dependent construction is necessary, given the {\em anisotropic} dispersions 
and {\em nonparallel} orientations of $Q$ valleys (Fig.~\ref{fig1}a).
In this representation, 
$H_f = (a^{\dagger}_f a_f + 1/2)\hbar\omega$ with $\omega=eB_{\perp}/\sqrt{m_xm_y}$.
The LL energies and wave functions are 
\begin{equation}\begin{aligned}
E_{fX}^{(n)} &= \left(n+{1}/{2}\right)\hbar\omega,\;\, n\in\{0,{\mathbb N}\},\;\, f\in\{u, d, s\},\\
\Psi_{fX}^{(n)} &= e^{i\bm{Q}_f\cdot\bm{r}}{L}_y^{-1/2}e^{ik_yy}
{N}_{fn}^{-1}e^{-\alpha_f\beta^*_f \xi_f^2/2}{H}_n(\xi_f),
\end{aligned}\label{Sol}\end{equation}
where $X=k_y\ell^2$ are the guiding centers, $\bm{Q}_f$ are the valley positions in the BZ, 
${L}_y$ is the system size along $\hat y$,
${N}_{fn} = (2^nn!\sqrt{\pi}|\alpha_f|\ell)^{1/2}$ are normalization factors, 
${H}_n(\xi_f)$ are Hermite polynomials, and $\xi_f = (x-X_f)/|\alpha_f|\ell$.      
The LLs of an odd-layer system is plotted in Fig.~\ref{fig1}c:
each level is triply degenerate due to the flavor symmetry, 
and quantum Hall effects occur at filling factors $\nu = 3N$ ($N\in{\mathbb N}$).
Note that the Zeeman splitting is larger than $\hbar\omega$ for $g^*m_{x,y} > 2 m_e$,
and that under an electric field the sextets in even-layer systems also split into triplets. 

{\color{blue}\indent{\it Flavor symmetry breaking.}}---Now we study the interaction effects  
at the integer filling factors $\nu = 3N-2$ and $3N-1$, {i.e.}, the highest LL triplet is $1/3$ filled or empty. 
We examine whether the emergent flavor symmetry and/or 
any other crystal symmetry is spontaneously broken and whether the broken symmetry states have  
charge excitation gaps that produce strong quantum Hall effects. 
Broken flavor symmetry ground states are either Ising-like or XY-like, 
as illustrated in Figs.~\ref{fig1}d-\ref{fig1}g.
For an Ising-like state, imbalanced flavor populations are energetically favored.
In the extreme limit, it can be a pure flavor state, {i.e.}, $|u\rangle, |d\rangle$, or $|s\rangle$. 
For an XY-like state, coherence among the flavors is spontaneously established. 
In the symmetric case, it can be a flavorless state such as $|u\rangle + |d\rangle + |s\rangle$. 
Markedly, the Ising-like states are nematics~\cite{Nematic,N1,N2,N3}, which break the $\mathcal{C}_3$ symmetry, 
whereas the XY-like states are incommensurate CDW states which lower the translational symmetry.  

We employ a Hartree-Fock theory~\cite{Cote,Exact,Girvin} to evaluate the interaction effects.  
The $\nu=3N-1$ and $3N-2$ states are particle-hole symmetric  
after projecting to the highest LL triplet. This allows us to focus on the $\nu=3N-2$ cases. 
We derive the mean-field Hamiltonians
\begin{align}
\begin{split}
H^{(n)}_{ff'} = \ & {E}^{(n)}\delta_{ff'} 
- \sum_{f''\neq f}{U}^{(n)}_{ff''}\Delta_{f''f''}^{(n)} \delta_{ff'} \\
&+ {W}^{(n)}_{ff'}\Delta_{f'f}^{(n)}(1-\delta_{ff'}) 
-{V}^{(n)}_{ff'}\Delta_{f'f}^{(n)},
\end{split}\label{HF}
\end{align}
where $n$ is the orbital index for the highest LL triplet,  
${E}^{(n)}$ has been given in Eq.~(\ref{Sol}), 
and $\Delta_{ff'}^{(n)}$ is the triplet density matrix that must be determined self-consistently. 
When projected to the subspace of the highest triplet,    
the flavor-conserving long-range interaction $V_{\bm k}={2\pi e^2}/{\epsilon k}$  
and the flavor-mixing short-range scattering $U_{Q}\sim{2\pi e^2 a}/{\epsilon}$ 
(with $\epsilon$ the dielectric constant) give rise to the Hartree matrix  
${W}_{ff'}^{(n)}=(2\pi\ell^2)^{-1}U_{Q}\,{F}_{ff'}^{nn}(0){F}_{f'f}^{nn}(0)$
and the two exchange (Fock) matrices in Eq.~(\ref{HF})
\begin{equation}\begin{aligned}
&{V}_{ff'}^{(n)} = \int\frac{d^2{\bm k}}{(2\pi)^2}\,V_{\bm k}\,
{F}_{ff}^{nn}({\bm k})\,{F}_{f'f'}^{nn}(-{\bm k})\,e^{i\varphi_{ff'}^{v}}, \\
&{U}_{ff'}^{(n)} = \int\frac{d^2{\bm k}}{(2\pi)^2}\,U_{Q}\,
{F}_{ff'}^{nn}({\bm k})\,{F}_{f'f}^{nn}(-{\bm k})\,e^{i\varphi_{ff'}^{u}}.
\end{aligned}\end{equation}
Here ${F}_{ff'}^{nn}$ is the generalized form factor~\cite{FF},
$\varphi_{ff'}^v = k_xk_y\ell^2(1-\varphi_{ff}-\varphi_{f'f'})$, 
$\varphi_{ff'}^u = k_xk_y\ell^2(1-\varphi_{ff'}-\varphi_{f'f})$ 
with $\varphi_{ff'} = \gamma_{f'}^*/(\gamma_f + \gamma_{f'}^*)$ 
and $\gamma_f = \beta_f/\alpha_f$. 
If the $Q$ valleys were isotropic, then $\alpha_f,\beta_f,\gamma_f=1$, 
$\varphi_{ff'}^{v,u}=0$, and ${F}_{ff'}^{00}=\exp{(-k^2\ell^2/4)}$, 
as in the case of circular cyclotron orbits~\cite{Cote}. 
Below, the flavor-dependent anisotropy plays an important role in determining the ground states.

The broken flavor symmetry ground state minimizes the total energy of Eq.~(\ref{HF}).
For the most general trial wave function parametrized as 
$(c_ue^{i\phi_u},c_de^{i\phi_d},c_s)^T$ with $\sum_fc_f^2=1$, 
the relative total energy per electron reads 
\begin{flalign}
{\bar E}_{T}^{(n)}=\left[{V}^{(n)}_S - {V}^{(n)}_D 
+{W}^{(n)}_D - {U}^{(n)}_D\right]\sum_{f\neq f'}c_f^2c_{f'}^2\,.
\label{Etot}
\end{flalign}
Because of the charge conservation of each flavor, 
${\bar E}_{T}^{(n)}$ does not depend on the phases $\phi_{u}$ and $\phi_{d}$;
yet ground states break these symmetries. 
The second factor in Eq.~(\ref{Etot}) is minimized by the three pure flavor states 
$|u\rangle=(1,0,0)^T$, $|d\rangle=(0,1,0)^T$, and $|s\rangle=(0,0,1)^T$, 
whereas it is maximized by those flavorless states $(e^{i\phi_u},e^{i\phi_d},1)^T/\sqrt{3}$. 
Thus, the sign of the first factor in Eq.~(\ref{Etot}) determines the nature of ground states.
Because of the flavor symmetry, the flavor-conserving exchange matrix has two inequivalent elements, 
the diagonal $V_S^{(n)}$ for electrons of the same flavor and the off-diagonal $V_D^{(n)}$ for electrons of different flavors. 
By contrast, the flavor-mixing matrices only have the off-diagonal elements 
$U_D^{(n)}$ and $W_D^{(n)}$ for electrons of different flavors.

Exchange energies are stronger between LL orbitals that are more similar. 
It follows that for the flavor-conserving interaction $V_S^{(n)}-V_D^{(n)}>0$ in Eq.~(\ref{Etot}), 
given the flavor-dependent anisotropy of Q valleys. 
This fact favors a nematic ground state $|u\rangle$, $|d\rangle$, or $|s\rangle$ that has a pure flavor, 
when the much weaker flavor-mixing scattering $U_Q$ can be ignored. 
While $V_S^{(n)}-V_D^{(n)}$ is positive and scales as $e^2/\ell\sim\sqrt{B_{\perp}}$, 
$W_D^{(n)}-U_D^{(n)}$ is negative and scales as $e^2a/\ell^2\sim{B_{\perp}}$. 
Despite $a<\ell=25.6/\sqrt{B_{\perp}[{\rm T}]}$~nm, at sufficiently large fields 
the short-range $U_Q$ dominates to favor 
the CDW states such as the flavorless $|u\rangle + |d\rangle + |s\rangle$.
Taking $U_{Q}={2\pi e^2 sa}/{\epsilon}$ ($\epsilon\simeq20$),
we find the first order transition at $B_{\perp}^c=1454/s^2$~T for the $n=0$ triplet. 

{\color{blue}\indent{\it Ferroelectricity.}}---For a pure flavor nematic state, 
all the electrons at Fermi energy are confined in one $Q$ valley. 
This valley is located at a non time-reversal-invariant momentum and lacks an inversion center.
For the flavor $u$ in Fig.~\ref{fig1}a, the parameter $\delta_x$ in Eq.~(\ref{Ham}) 
characterizes the band distortion along $\hat x$.  
In light of a recent theory on quantum Hall ferroelectrics~\cite{F00}, 
the state $|u\rangle$ must carry an electric dipole moment along $\hat y$.
In the Berry phase approach to polarization~\cite{P1,P2,P3,P4}, the electric dipole moment per electron reads
\begin{equation}
p_u^{(n)} = -\frac{e}{L_x}\int_{0}^{L_x} \Big\langle u_{uX}^{(n)} \Big| i\nabla _{k_y} \Big| u_{uX}^{(n)} \Big\rangle dX\,,
\label{BP1}\end{equation}
where $L_x$ is the system size along $\hat x$, 
and $| u_{uX}^{(n)} \rangle$ is the cell periodic eigenstate for $k_y=X/\ell^2$. 
By applying the $x-k_y\ell^2$ dependence of $u_{uX}^{(n)}(x)$ to Eq.~(\ref{BP1}), we obtain 
\begin{equation}
p_u^{(n)} =e\ell^2\int_{-\infty}^{+\infty} u_{u0}^{(n)^*}(x) i\nabla _{x} u_{u0}^{(n)}(x) dx\,.
\label{BP2}\end{equation}

\begin{figure}[t!]
\includegraphics[width=1.0\columnwidth]{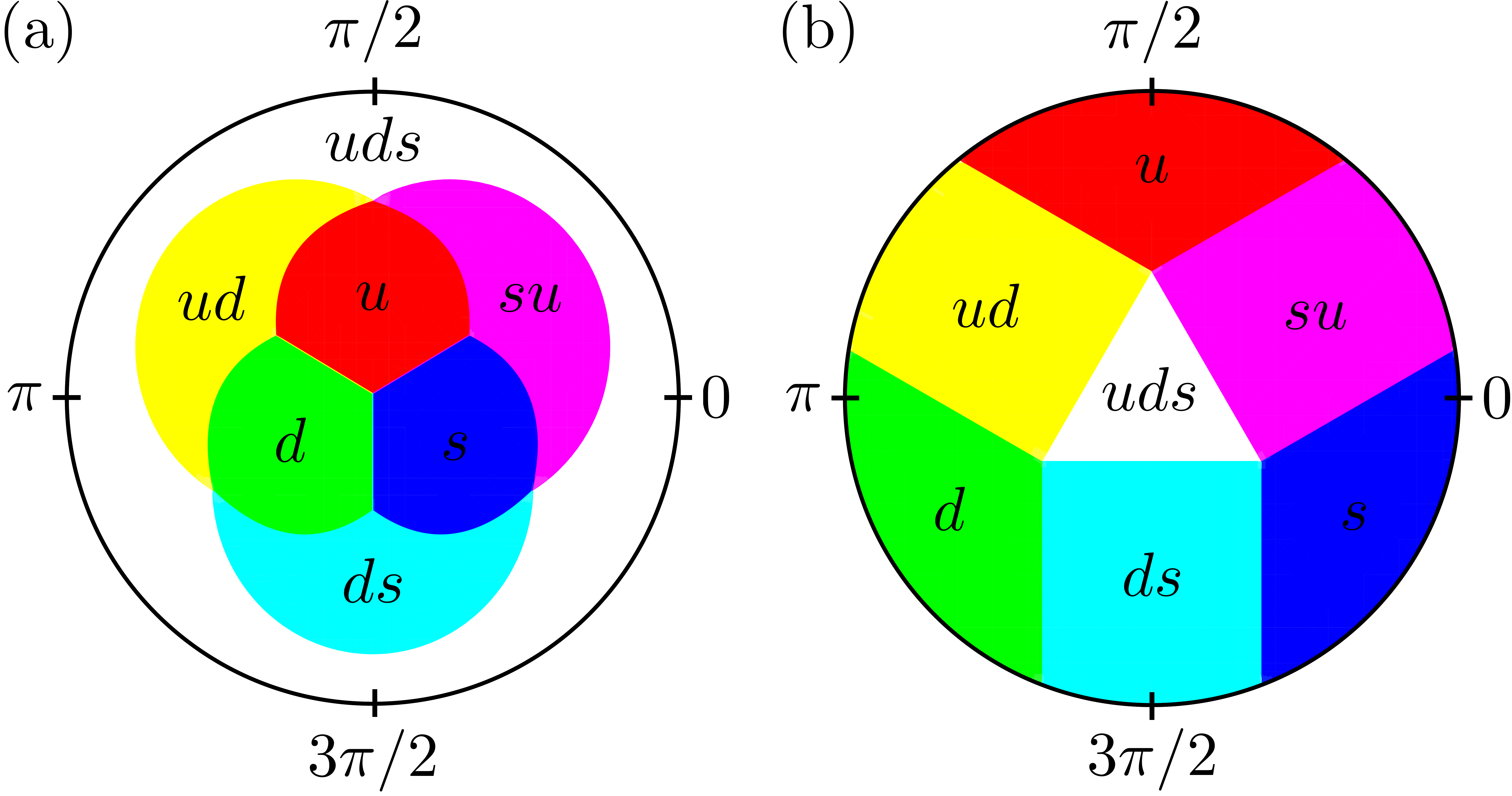}
\caption{Phase diagrams at $\nu=1$ versus the field strengths (radial) and the orientation of $E_{\parallel}$ (angular).
(a) $B_{\perp}$ varies from $0$ to $40$~T at $E_{\parallel} = 1.5$~mV$\cdot$nm$^{-1}$. 
(b) $E_{\parallel}$ varies from $0$ to $2.0$~mV$\cdot$nm$^{-1}$ at $B_{\perp} = 20$~T.
$f\in \{u, d, s\}$ label the pure flavor ferroelectric nematics; 
$ff'$ and $uds$ label the two- and three-flavor coherent states with generally unequal flavor populations. 
The lines at $\theta=3\pi/2-\theta_f$ in (a) are first order transitions between different pure flavor states; 
all other lines are continuous transitions between states with coherence among different number of flavors. 
We have chosen $s=10$.}
\label{fig2}
\end{figure}

For the undistorted case, 
$u_{u0}^{(n)}(x)$ is the wave function in Eq.~(\ref{Sol}) but without the first three factors.
Because of the definite parity of $u_{u0}^{(n)}(x)$, Eq.~(\ref{BP2}) vanishes as expected.
For the distorted case, by employing the perturbation theory to the $2M$-th order, we find 
\begin{equation}
p_u^{(n)} = \sum_{m=1}^{M}C_{2m-1} \left(\frac{\delta_x a}{\ell}\right)^{2m-1} e\ell\,,
\label{dm}\end{equation}
where ${C}_{2m-1}$ are dimensionless coefficients and ${C}_{2m}=0$. Evidently, the dipole moment per electron scales 
with $B_\perp^{m-1}$ perturbatively at the $(2m-1)$-th order; 
the leading order does not depend on $B_\perp$ and must survive even at $B_\perp=0$.
As the distortion has larger effects at higher energies, 
the higher the LL index $n$, the larger the $p_u^{(n)}$. 
For the $n=0$ triplet, we obtain at the leading order
\begin{equation}
p_u^{(0)}=\frac{m_x}{m_y}\,\frac{\sqrt{m_xm_y}}{m_e}\,\frac{3\delta_x}{2}\,ea\,.
\end{equation}

This produces a dipole moment $0.04ea$ per electron for trilayer MoS$_2$, 
which amounts to a polarization density $5.0B_{\perp}[{\rm T}]\times10^{-4}$~pC$\cdot$m$^{-1}$,
given that the electron density per Landau level is $1/2\pi\ell^2$. 
For comparison, the recently discovered 2D ferroelectric materials have polarization densities~\cite{F1,F2,F3,F4} 
of $10-100$~pC$\cdot$m$^{-1}$. Whereas these 2D materials are structural ferroelectrics, 
the ferroelectricity of our quantum Hall nematics is an electronic property emergent from 
spontaneous symmetry breaking in a dilute electron gas, {i.e.}, 
$\sim1$ electron per $2000$ sites at $B=20$~T (or $a^2/2\pi\ell^2$). 
In this sense, our obtained dipole moment per electron is reasonable and large.

{\color{blue}\indent{\it Electric field effect.}}---As a result, 
an electron in the $n$-th LL of flavor $f$ carries an in-plane dipole moment 
$p_f^{(n)}=p_u^{(n)}$ at an angle of $\pi/2+\theta_f$ from the $x$-axis. 
This implies that an in-plane electric field $E_{\parallel}$ couples to 
the nematic order parameters by breaking the $\mathcal{C}_3$ symmetry 
and influences the flavor ordering by producing 
flavor-dependent electrostatic energies
\begin{equation}
{\delta E}^{(n)}_f = - p_f^{(n)}E_{\parallel} \cos\left(\theta_f + \frac{\pi}{2} - \theta\right),
\label{Ec}\end{equation}
where $\theta$ is the in-plane orientation angle of $E_{\parallel}$ from the $x$-axis.
It follows that ${E}^{(n)}\rightarrow{E}^{(n)}+{\delta E}^{(n)}_f$ in Eq.~(\ref{HF}). 
In the absence of interactions, the $E_{\parallel}$ field lifts the triplet degeneracies, 
favors pure flavor states, and creates integer quantum Hall effects 
at $\nu=3N-2$ and $3N-1$ except for several special orientations. 
For $\theta=3\pi/2-\theta_f$, there is still a twofold degeneracy at $\nu=3N-2$, 
and likewise for $\theta=\pi/2+\theta_f$ at $\nu=3N-1$.

The electrostatic energies compete with the electron-electron interactions by adding 
$\sum_fc_f^2{\delta E}^{(n)}_f$ to Eq.~(\ref{Etot}). This greatly enriches the phase diagrams.  
Figure~\ref{fig2}a displays a typical phase diagram at $\nu = 1$ ($n=0$) 
for $E_{\parallel} = 1.5$~mV$\cdot$nm$^{-1}$. Below the $B_{\perp}^c$ derived above,
interactions favor a pure flavor nematic state, and $\theta$ selects the flavor,
with first order transition lines at $\theta=3\pi/2-\theta_f$. 
Above $B_{\perp}^c$, as interactions prefer to mix flavors, 
and electrostatic energies stabilize two-flavor and three-flavor coherent states 
with generally unequal flavor populations that are continuously tunable by the two fields. 
These features are also exhibited by Fig.~\ref{fig2}b at a fixed $B_{\perp}>B_{\perp}^c$. 
If $E_{\parallel}$ is smaller in Fig.~\ref{fig2}a, 
the critical points at $3\pi/2-\theta_f$ remain $B_{\perp}^c$ whereas those
at $\pi/2+\theta_f$ move closer to $B_{\perp}^c$,  
and the space for the two-flavor states shrinks.
Note that $B_{\perp}^c$ would be enhanced for larger $Q$-valley anisotropy or 
weaker short-range interactions.

{\color{blue}\indent{\it Discussion.}}---In quantum transport,  
Hall plateaus and SdH minima at $\nu=3N\pm1$ can be strong evidence for the flavor symmetry breaking. 
The nematic states can be imaged by Scanning Tunneling Spectroscopy via impurity scattering~\cite{BiLL} 
and probed by measuring the anisotropy in optical conductivity or dielectric function.
All these signatures are highly sensitive to the external fields $E_{\parallel}(\theta)$ and $B_{\perp}$. 
It would also be intriguing to explore the Pockels and Kerr electro-optic effects in birefringence,   
utilize the hysteresis of nematic ferroelectricity for tunable capacitor and memory functions, 
and examine the effects of uniaxial strain that breaks the flavor symmetry.

Pure flavor nematic states can be pinned by atomic impurities. 
As domain walls proliferate thermally above a critical temperature~\cite{DW,Shayegan,Abanin,Kumar},
the temperature to which the nematics survive is limited by domain wall nucleation. 
In sharp contrast to many other nematics~\cite{Nematic,N1,N2,N3}, 
the ones studied here have full bulk gaps and chiral edge states. 
Thus, the domain walls host counter-propagate edge states when flavor-mixing scattering 
is negligibly weak~\cite{DW6,DW1,DW2,DW3,DW4,DW5}. 
Moreover, the emergent ferroelectricity enables manipulation of the domain walls  
by external fields such as $E_{\parallel}(\theta)$ and $B_{\perp}$, as showcased in Fig.~\ref{fig2}.

The quantum Hall nematicity and electric polarization density are continuously 
tunable by $E_{\parallel}(\theta)$ and $B_{\perp}$ fields (Fig.~\ref{fig2}).
The dipole moment of a given flavor is reversed by reversing the $B_{\perp}$ field.  
Yet, this also switches from the $Q$ to $Q'$ valleys (Fig.~\ref{fig1}a) 
that have opposite spin splittings and band distortions.
Thus, the measured polarization is not expected to reverse. 
Finally, the fact that the leading contribution to the dipole moment per electron in Eq.~(\ref{dm}) 
is independent of $B_{\perp}$ implies the existence of flavor-dependent dipole moments at $B_{\perp}=0$. 
Therefore, while the out-of-plane orientation of Zeeman field selects the locked spin and valley, 
the in-plane orientation of electric field selects the flavor of electrons, 
giving rise to a new concept: the flavortronics.  

{\color{blue}\indent{\it Acknowledgement.}}---We thank Philip Kim, Allan MacDonald, Kin Fai Mak, Joe Qiu, Ning Wang, 
and Emanuel Tutuc for helpful discussions. We are grateful to Xiao Li and Gui-Bin Liu for technical help.


\begin{thebibliography}{10}
\bibitem{LLnote} Fig.~\ref{fig1}c plots the LLs of the lowest spin-split subbands. 
The Zeeman splitting is between the opposite spins respectively in the $Q$ and $Q'$ valleys. 
The LL structure for the next spin-split subbands is almost identical but at a few tens of meV higher in energy, 
producing LL crossings.
\end{thebibliography}
\end{document}